# Measurement of the Cotton-Mouton effect in nitrogen, oxygen, carbon dioxide, argon, and krypton with the Q & A apparatus

Hsien-Hao Mei [†], Wei-Tou Ni [†,*], Sheng-Jui Chen [‡], and Sheau-shi Pan [‡]

(Q & A Collaboration)

[†] Center for Gravitation and Cosmology, Department of Physics, National Tsing Hua University, Hsinchu, Taiwan 30013, Republic of China
[‡] Center for Measurement Standards, Industrial Technology Research Institute, Hsinchu, Taiwan 30015 Republic of China

**Abstract**

Experiments for vacuum birefringence and vacuum dichroism have been set up with high-finesse high magnetic experimental apparatuses, which seem to be ideal for small gaseous Cotton-Mouton effect (CME) measurements. PVLAS Collaboration has measured CMEs in krypton, xenon and neon at the wavelength of 1064 nm. In this Letter, we report on our measurement of CMEs in nitrogen, oxygen, carbon dioxide, argon, and krypton at the same wavelength in a magnetic field $B = 2.3$ T at pressure $P = 0.5$-300 Torr and temperature $T = 295$-298 K. Our results agree with the PVLAS results in the common cases.

## 1. Introduction

Upon passing through a medium with transverse magnetic field, linearly polarized light becomes elliptical polarized. Cotton and Mouton [1] first investigated this in detail in 1905, and the phenomenon is known as Cotton-Mouton effect (CME). Systematic experimental work of Buckingham et al. in 1967 [2] was followed by many publications of different groups [3-9] on gases up to 1991. In 1997, Rizzo et al. [10] made a systematic comparison between measurements and theoretical models.

According to QED (Quantum Electrodynamics), vacuum also has such a birefringence [11-12]. PVLAS [13] Collaboration and Q & A Collaboration [14] are specifically looking for such a QED vacuum birefringence. With high requirements on sensitivity, the apparatuses of these experiments are applicable to the gaseous CME measurement, especially for tiny birefringent effects, such as that of hydrogen, neon, helium, or gas under low pressure or in low magnetic field. PVLAS Collaboration has recently measured Cotton-Mouton effects in krypton, xenon, nitrogen [15], neon [16], with an earlier result on oxygen [17] at the wavelength of 1064 nm. In this Letter, we report our results at the same wavelength for various gases. The results on argon and carbon dioxide are new.

The small ellipticity induced in gas can be measured from a pair of crossed polarizers with high extinction ratio. The magnetic birefringence is due to difference in the index of refraction $\Delta n$ for light polarization parallel and perpendicular to the external transverse magnetic field **B** and is proportional to the magnitude of this field $B$ to the second power. For dilute gas, $\Delta n$ is proportional to the pressure $P$. For the Cotton-Mouton coefficient we will follow [10] and use the normalized Cotton-Mouton birefringence $\Delta n_u$ at $P = 1$ atm and $B = 1$ T. With this definition, we have

$$\Delta n \equiv n_{\parallel} - n_{\perp} = \Delta n_u \, (B/1 \text{ T})^2 \, (P/1 \text{ atm}). \qquad (1)$$

The phase retardation $\delta$ of light between these two polarization directions is:

$$\delta = (2\pi/\lambda)\Delta n L_B, \qquad (2)$$

where $L_B$ is the effective length of the magnetic field, and $\lambda$ the wavelength of light. For incident light with polarization making an angle $\theta$ with respect to the transverse magnetic field **B**, the ellipticity for the exiting light is $(\delta/2)\sin 2\theta$. For a Fabry-Perot cavity with finesse $F$, light in average travels $2F/\pi$ times before exiting the cavity; hence the retardation in (2) is enhanced by $2F/\pi$, and the acquired ellipticity $\psi$ is

$$\psi = (2F/\pi)(\delta/2)\sin 2\theta. \qquad (3)$$

For a dipole permanent magnet rotating with the circular frequency $\omega_m$ (= $2\pi f_m$), the angle $\theta(t)$ is $\omega_m t$. Denote $\Psi$ to be the Fourier spectrum of $\psi(t)$. From (1) - (3), the amplitude of $\Psi$ at $2\omega_m$, $\Psi_0$ [$\equiv \Psi(2\omega_m)$], is

$$\Psi_0 = \Delta n_u (2F/\lambda)\{\int (B/1\text{T})^2 \, \mathrm{d}L\}(P/1 \text{ atm}) + b$$
$$= aP + b, \qquad (4)$$

where $b$ is the constant part of the ellipticity which comes from imperfection of apparatus and vacuum birefringence. The coefficient $b$ is small and negligible compared to uncertainties in this paper. Nevertheless, we keep it to monitor errors. Let $\{\int B^2 \, dL\}$ be denoted as $\kappa$, from (4), $\Delta n_u$ is given by

$$\Delta n_u = a \, (2\kappa F/\lambda)^{-1} \cdot (1 \text{ T})^2 \cdot (1 \text{ atm}). \qquad (5)$$

The experimental value and uncertainty for slope $a$ are determined by fitting to the ellipticity measurement data, while that for $F$ and $\kappa$ are determined by separate finesse and magnetic-field measurements. The statistical uncertainty for $\Delta n_u$ is

$$\sigma_{\Delta nu\text{-}stat} = \Delta n_u \, [(\sigma_a/a)^2 + (\sigma_F/F)^2]^{1/2}, \qquad (6)$$

and the systematic uncertainty for $\Delta n_u$ is given by the calibration uncertainties for $P$ and $B$, and the modulation depth ($\eta_0$) uncertainty of the Faraday Cell modulator (a normalization factor to be explained in section 2):

$$\sigma_{\Delta nu\text{-}sys} = \Delta n_u \, [(\sigma_P/P)^2 + (2\sigma_B/B)^2 + (\sigma_{\eta 0}/\eta_0)^2]^{1/2}. \qquad (7)$$

We started to build the Q & A experiment in 1994 to test the Quantum electrodynamics (QED) and to search for Axion by ultra sensitive interferometer [18] using the noise-isolation techniques developed in the Gravitational Wave Detection Community [19]. In 2003, we completed a prototype, and measured CME and Verdet effect of the air at 1 atm [20]. During the second stage, we have used polarizers of high extinction ratio (up to 92.87±0.11 dB) for ellipsometry [21] and effective close-loop control of the mirror suspension system [22] to make a 19.2 hours integration of vacuum dichroism measurement giving an upper limit [14] on pseudoscalar-photon interactions [23]. We have also measured the CME of $N_2$ at pressure range from 2-150 Torr [14, 24].

In this Letter, we report on novel measurements of the CMEs of $CO_2$ and Ar at laser wavelength of 1064 nm ($T = 295$-298 K). The measurements of the CMEs for $N_2$, $O_2$, and Kr are made for reference and calibration.

## 2. Experiment setup

The Q & A experiment apparatus, as shown in Fig. 1, is discussed in detail in [14,21,22,24,25]. For incoming linearly polarized light gone through the center of 27 mm borehole of the dipole magnet, the magnetic field rotating at $f_m = \omega_m/2\pi = 6.7742$ rev/s generates a modulated ellipticity signal at $2f_m$ due to the Cotton-Mouton birefringence. After transmitting through a quarter-wave plate (QWP), this signal becomes a modulated ($2f_m$) polarization rotation signal. The QWP is adjusted slightly in addition to compensate the small cavity mirror (dc) birefringence $\zeta$ with an angle $\xi$ between its slow axis and incident polarization $\mathbf{E}$ for improving the performance of system's total extinction ratio and achieving a better angular resolution. The light from the QWP is further modulated by the Faraday Cell (FC) with polarization-rotation modulation depth $\eta_0 = 837 \pm 13$ μrad at frequency $f_f = \omega_f/2\pi = 390$ Hz. The effect is to shift the original ellipticity signal at $2f_m$ to higher frequency side-bands ($f_f \pm 2f_m$) for easier detection. After analyzed by an analyzer (A) and detected by a photo diode (PD2 to Ch1 [Channel 1]), the signal is demodulated at $\omega_f$ (in-phase amplitude $X_1$ at $\omega_f$ to Ch2; quadrature-phase amplitude $Y_1$ at $\omega_f$ to Ch3) and $2\omega_f$ (in-phase amplitude $X_2$ at $2\omega_f$ to Ch4; quadrature-phase amplitude $Y_2$ at $2\omega_f$ to Ch5) by two Lock-in Amplifiers (LA) and is recorded. Cotton-Mouton birefringence shows up at the $2\omega_m$ component of the in-phase amplitude $X_1$.

We have calculated in detail in [25] all terms of the demodulated signal from the lock-in amplifiers including (i) the ellipticity of gaseous CME and/or vacuum birefringence $\Psi$ generated by the transverse magnetic field, (ii) ellipticity of mirror birefringence $\zeta$ generated by stray magnetic field, (iii) the polarization rotation due to vacuum dichroism $\beta$ generated by the transverse magnetic field, and (iv) the gaseous/mirror Verdet effect $\upsilon/\upsilon_M$ generated by small residual/stray axial magnetic field. The major terms relevant to this investigation are listed in Table 1. For the CME measurement, $\beta << \Psi_0$, $\zeta << \Psi_0$, $\upsilon << \Psi_0$, and $\upsilon_M << \Psi_0$. The demodulated signal $X_1$ is reduced to $\eta_0 I_0(t) \Psi_0 \sin 2\omega_m t \, [= \eta_0 I_0(t) \psi(t)]$, where $I_0(t)$ is the light intensity. Dividing $X_1(t)$ by $Y_2(t)$ to take out the $I_0(t)$ dependence, we obtain the time-domain ellipticity signal:

$$\psi(t) \equiv (\eta_0/4)[X_1(t)/Y_2(t)]. \qquad (8)$$

The gaseous ellipticity signal $\Psi_0$ is thus experimentally given by the $2\omega_m$ amplitude of the spectrum of $\psi(t)$ as defined just before Eq. (4). The modulation depth $\eta_0$ is a normalization factor. Its experimental error gives a contribution to the systematic uncertainty in Eq. (7) for the birefringence measurement.

During the whole measuring process, the frequency of 1064 nm Nd-YAG laser is locked to the 3.45 m long high-finesse Fabry-Perot interferometer (FPI) by Pound-Drever-Hall technique. The Fabry-Perot mirrors $CM_1$ and $CM_2$ are suspended by two sets of X-pendulum-double-pendulum system for vibration isolation. The finesse of the FPI is measured to be $F = 30,000 \pm 1,673$ using a ringing model [26] (Fig. 2).

The measurements of transverse magnetic field vs. position of the permanent magnet in December 2002 [27,28] and November 2008 are shown in Fig. 3. The maximum value after manufacture is 2.3 T. The integrated value of $\kappa$ ($= \int B^2 \, dL$) is 2.532 $T^2$m in November, 2008. The accuracy of the Hall probe in this measurement is 1.1%, therefore the calibration uncertainty for $\kappa$ is 2.2%.

We use Ch6 to record the signals of a photodiode used to detect the reflection light intensity of a special mirror on the rim of the rotating magnet by a PC at a sampling rate 20 kHz, together with other signals in Ch2 - Ch5. The start time of each rotation cycle is determined from the rising curve of received laser light intensity from the special mirror. A fixed threshold is used for this determination by data-point interpolation.

Two vacuum chambers housing the two Fabry-Perot mirrors with a total volume about 3000 liters are pumped by a turbo molecular pump and a mechanical pump. The gas-line to gas storage bottles is controlled by several angle valves. The gaseous pressure is recorded (Ch7) by two gauges with a capacitance sensor and piezo-resistor sensor respectively. The readings are interpolating-corrected under 95% confidence level according to the Student's t-distribution using calibration tables made for the two gauges. These calibration tables have been traced back to NML (National Measurement Laboratory, R.O.C.) calibration standards C940810 and C940811, and German National Laboratory PTB (Physikalisch-Technische Bundesanstalt) calibration standards 2233PTB07 and 2232PTB07. The calibration error for $P$ in the region of our measurements ranges from 1.77% to 2.48%.

## 3. Results and discussions

All useful signals were acquired at a sampling rate of 20 kHz by a PC-based data acquisition board (DAQ) using the PC internal clock, and saved to a hard disk for later off-line software lock-in analysis with respect to the phase of the magnet rotation ($f_m \sim$ 6.7742 cycle/s). Synchronization processing with the magnet rotation is achieved by off-line interpolations of the phase-marked data (Ch6) for each rotation cycle. We use the determined start time of each rotation cycle to divide the cycle into 128 equal parts by interpolation method, and use them to produce a new set of data (128 points) from the PC acquired 2952-2954 data points within each rotating cycle of each channel (Ch2 - Ch5). From these re-sampled data at sampling rate of $128f_m$, we calculate the elliptic signal $\psi(t)$ from Eq. (8). Spectra of $\psi(t)$ are taken for every consecutive 256 magnet rotating cycles. The gaseous ellipticity signal $\Psi_0$ is derived from the signal amplitude at $2f_m$ of the re-sampled spectra. These signals are then vector averaged to obtain measured values and standard deviations.

We measured the ellipticity due to gaseous CMEs of $N_2$, $O_2$, $CO_2$, Ar, and Kr at 5-6 different pressures from 0.5 to 300 Torr. The purity of these gases was better than 99.99%, and the room temperature was kept in the range of 294.86-297.96 K. At each pressure set point for a specific gas, the data-taking was no less than 3 hours. Over 73,000 magnetic rotating cycles (280 spectra) were recorded and vector-averaged at each pressure set point for all gases. The signal to noise ratio for all spectra were better than 100, even for the smallest detected ellipticity signal of Kr at 30 Torr in the measurement.

In Fig. 4 we show the vector-averaged ellipticity signals at frequency $2f_m$ from the spectra of the gaseous CMEs of $CO_2$ and Ar at laser wavelength 1064 nm. The positions of the centers of small circles are the averaged ellipticities, and the radii of them are twice the standard deviation ($2\sigma$). The phases are defined with respect to magnet rotation. The phase delay ($23.13°$) of the notch filters and low pass filters of the lock-in amplifiers is estimated from the data and subtracted. The phase difference ($\sim 180°$[179.854°±0.160°]) between diagrams of $CO_2$ and Ar implies a sign change of $\Delta n_u$ defined in Eq. (1). The ellipticity vs. pressure diagrams for $N_2$, $O_2$, $CO_2$, Ar, and Kr are shown in Fig. 5. The $\chi^2$ fitting results of the linear model in Eq. (4), where the reciprocal square of errors are used as weight for average, are shown in the legend. Statistical pressure uncertainties and ellipticity measurement uncertainties are included in the error-budget. The solid line is the fitted curve.

A comparison of $\Delta n_u$ measurements of $N_2$, $O_2$, $CO_2$, Ar, and Kr reported since 1905 to those from this work is compiled in Table 2. The statistical uncertainties $\sigma_{\Delta nu\text{-}stat}$ for present work are calculated from Eq. (6), and the systematic errors $\sigma_{\Delta nu\text{-}sys}$ are calculated from Eq. (7). Systematic uncertainties due to $P$, $B$ and $\eta_0$ are no larger than 3.66%, and those due to environmental temperature fluctuation, degassing, or impurity are relatively small. The overall rss (root-sum-square) systematic error $\sigma_{\Delta nu\text{-}sys}$ is no larger than 3.86%, which is adopted universally in the presentation of our results. The total uncertainty is the root-sum-square of the statistic uncertainty and the systematic uncertainty.

For the CMEs of krypton and nitrogen at 1064 nm, both PVLAS Collaboration [15] and our team (Q & A Collaboration) have performed measurement and the results are in agreement within experimental uncertainty (1 $\sigma$). Our result for oxygen agrees with an earlier PVLAS measurement [17] to within 1.2 $\sigma$.

Current measurement of $\Delta n_u$ for nitrogen agrees with our former result in nitrogen [6] to within 1.1 $\sigma$. In this former measurement of nitrogen, we use a different set of cavity mirrors with cavity finesse 4,400. The large systematic uncertainty comes from the 15% accuracy of the pressure gauge as stated by manufacturer.

For the present measurement, the electronic fluctuations of pressure readings dominate the statistical uncertainty $\sigma_P/a$ in the pressure range between 10 to 100 Torr. The pressure readings were obtained from two different pressure gauges for different pressure ranges of interest. The least accurate part of the full scale 1000 Torr gauge (Hastings HPM-2002-OBE) was adopted for the connecting region of 10-100 Torr. For improvement, more accurate and stable pressure gauges will be acquired and used for our future work.

## 4. Outlook

Axial molecular CME model for $N_2$, $O_2$ and $CO_2$, was quite different theoretically [33] from that of noble gases such as Ar and Kr. Some nonlinear dependence in powers of pressure $P$ [32] and the inverse absolute temperature $T$ [7] was expected. A better environmental control and more precise CME measurements will be able to test these theoretical models. We look positively on the potential of our apparatus for the CME measurements in gases. Moreover, this can always provide calibration for our vacuum birefringence measurement.

Upgrades are undergoing for a new 2.3 T dipole permanent magnet with length $L_B = 1.8$ m, a new laser system with optics of wavelength 532 nm, and a pair of new Fabry-Perot cavity mirrors with finesse 100,000. The ellipticity signal strength will be enhanced and the sensitivity to $\Delta n$ will be improved in these upgrades by a total factor of 20. This together with better calibrations of pressure gauges and magnetic probes and better control in environmental temperature will improve the precision and accuracy of our CME measurements.

**Acknowledgements**

We thank the National Science Council (NSC 96-2119-M-007-004 and NSC 97-2112-M-007-002) for supporting this program, and Prof. Jow-Tsong Shy for lending equipments.

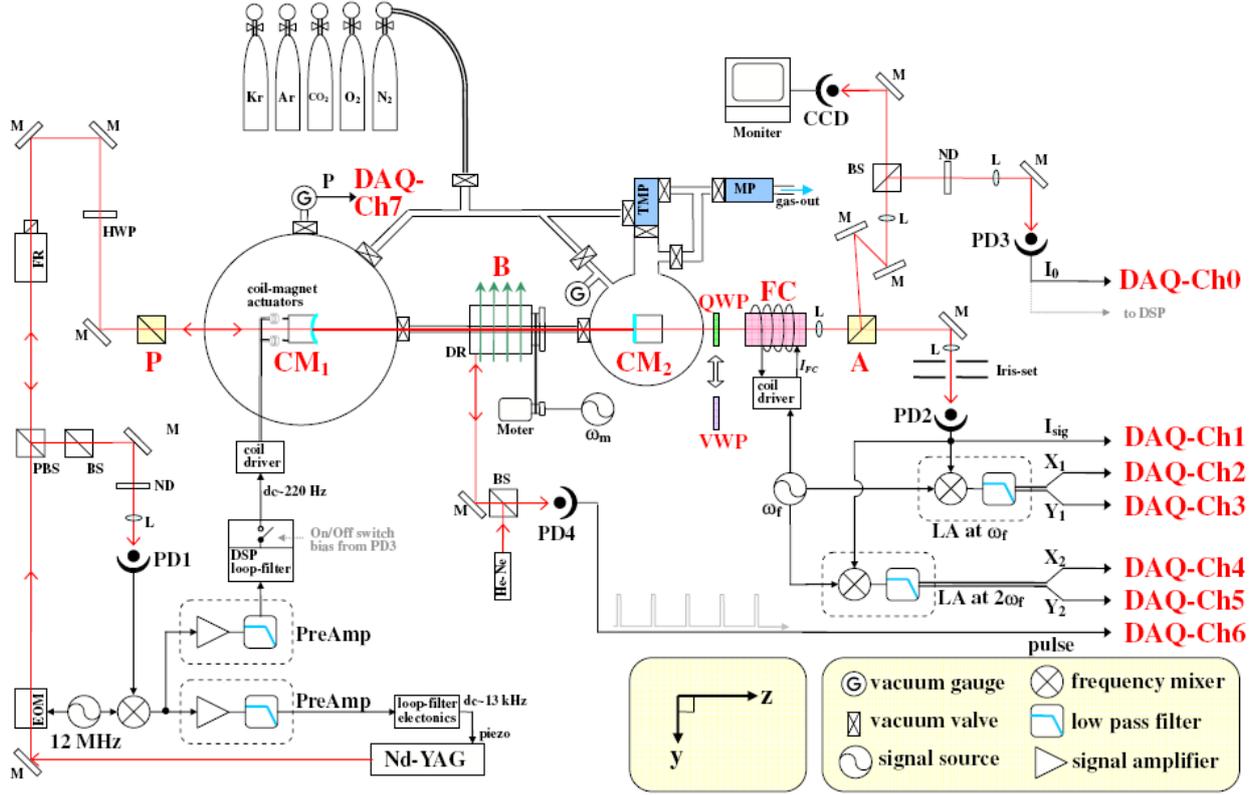

Fig. 1. Q & A experimental setup. The Nd-YAG laser frequency and the Fabry-Perot cavity (CM₁, CM₂) length are mutually locked to each other through the Pound-Drever-Hall technique. One 0.6 m 2.3 T dipole permanent magnet, rotating with angular speed $\omega_m$, generates the birefringence signal of the gas inside the chamber. Small ellipticity is accumulated after the light bouncing back and forth in the cavity mirrors, and converted into polarization rotation through a quarter wave plate (QWP). A coil-driving Faraday glass (Faraday cell, FC) provides polarization rotation carrier frequency $\omega_f$ so that the birefringence signal is moved into the side-bands. Light is extinguished by the polarizer A; the AC birefringence signal of the gas can be detect through a photo-diode (PD2). Lock-in detections are implemented both online with hardware (LAs with $\omega_f$ and $2\omega_f$) and offline with software (interpolation to Ch2-Ch5 signals in synchronization to the $\omega_m$ triggers from PD4).

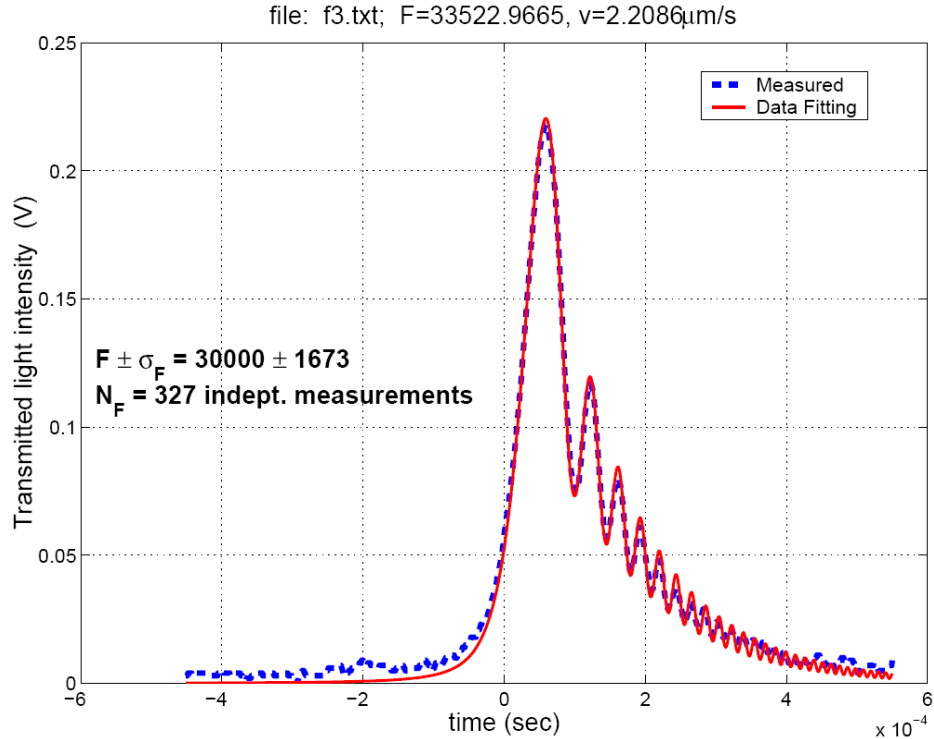

Fig. 2. An example of the finesse measurement using a ringing down Airy model [26] for fitting. The solid line is the fitted curve. In this example, the cavity mirrors moved with relative velocity 2.2086 μm/s and swept over one resonance at time 0. The finesse in this example is 33,523.

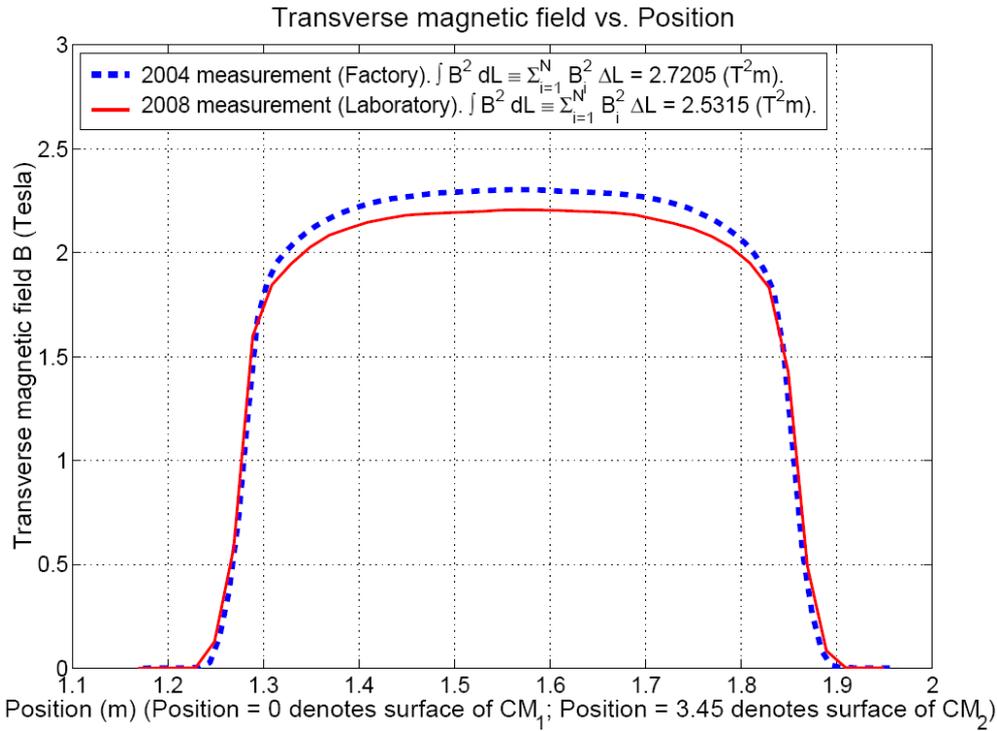

Fig. 3. The transverse magnetic field vs. position of the rotating magnet in 2008 (solid), and comparison of the field measured in 2004 (dashed). A 3.54% decay of the field is observed.

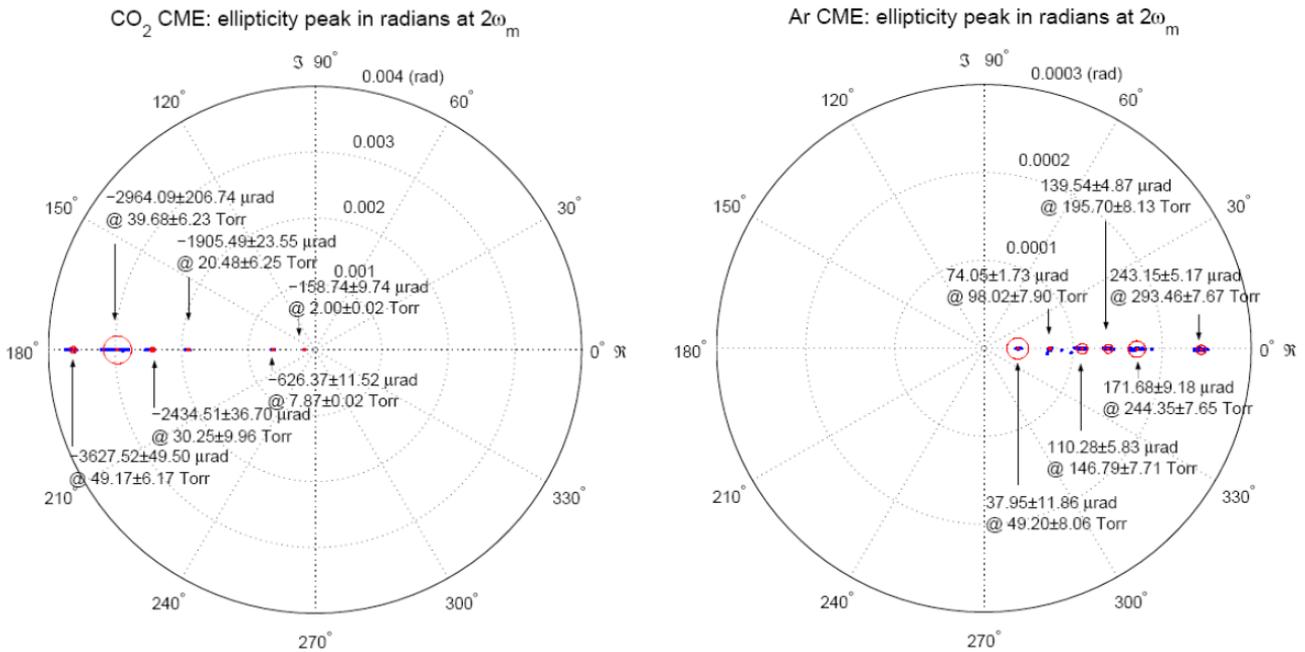

Fig. 4. The vector-averaged ellipticity signal from twice the rotation frequency of the spectrum of the gaseous Cotton-Mouton effect of $CO_2$ and Ar in 1064 nm. The phase delays due to the notch filters and low pass filters of the lock-in amplifiers are only approximately known to be around 30°; a constant phase 23.13° is fitted to the data for the resulting birefringence to be on the real axis. The phase difference (~180°) between diagrams of $CO_2$ and Ar implies a sign change of $\Delta n_u$.

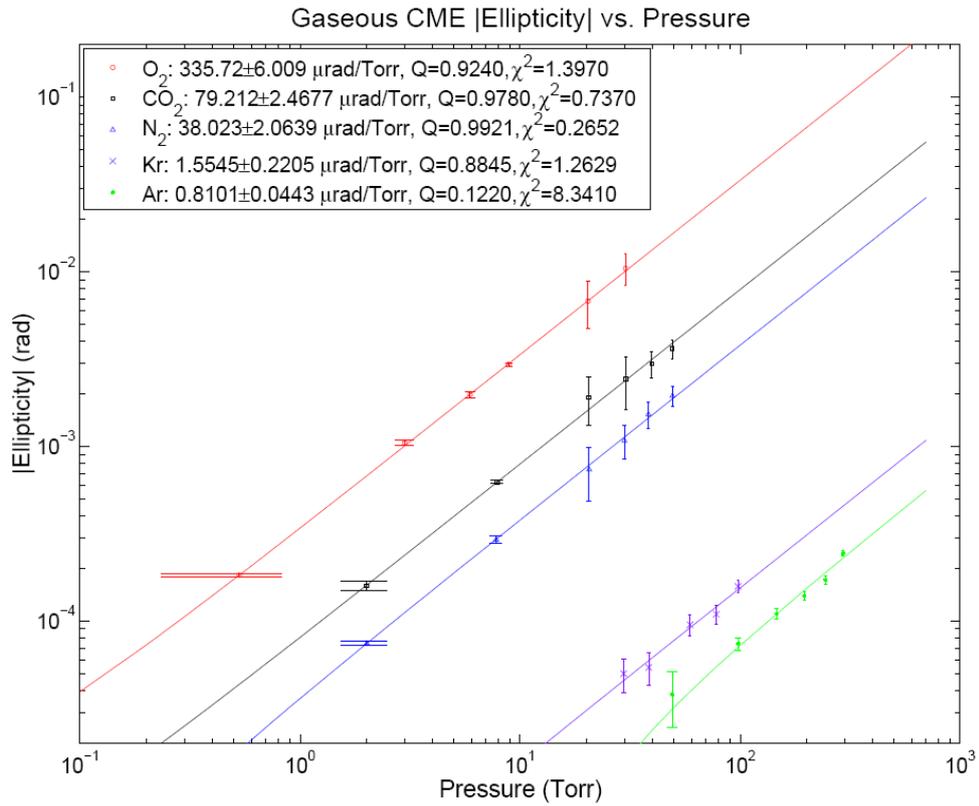

Fig. 5. Gaseous CME |ellipticity| vs. pressure. The ellipticities are regarded to be linear to pressure. The statistical pressure uncertainties are embedded in the error bars of that in the ellipticity signal.

Table 1
Major terms of demodulated signal from the lock-in amplifiers.

| Ch. | Major components of the channel |
|---|---|
| $X_1$ | $\eta_0\, I_0(t)\,[\,\Psi_0\sin2\omega_m t - \beta\zeta\sin(2\omega_m t+2\xi) - \zeta\cos2\xi\,(\upsilon+\upsilon_M)\sin(\omega_m t+\phi) + (\zeta\Psi_0/2)\,(\upsilon+\upsilon_M)\sin(\omega_m t-\phi)\,]$ |
| $Y_1$ | 0 |
| $X_2$ | 0 |
| $Y_2$ | $(\eta_0/2)^2\, I_0(t)$ |

$I_0(t)$: intensity detected on photo-diode PD2 and amplified by Lock-in Amplifiers (LAs); $\eta_0$: modulation depth of FC; $\Psi_0$: ellipticity of gaseous CME/vacuum birefringence; $\zeta$: ellipticity of mirror birefringence; $\xi$: angle between incident polarization and the slow axis of mirror birefringence; $\beta$: vacuum dichroism; $\upsilon/\upsilon_M$: gaseous/mirror Verdet effect; $\phi$: phase of residual axial magnetic field with respect to the transverse magnetic field modulation. Since $\beta\ll\Psi_0$, $\zeta\ll\Psi_0$, $\upsilon\ll\Psi_0$, and $\upsilon_M\ll\Psi_0$, they can be neglect in CME measurement; therefore $X_1\approx\eta_0\,I_0(t)\,\Psi_0\sin2\omega_m t$.

Table 2
Comparison of experimental data for different gases of $\Delta n_u$.

| Gas | Ref. | $(\Delta n_u\pm\sigma_{\Delta nu})\times10^M$ | $\lambda$(nm) | $T$(K) | $P$(Torr) |
|---|---|---|---|---|---|
| $N_2$ | [2] | -2.47±0.17 | 546.1 | 193.15-293.15 | 760 |
| (M = 13) | [3] | -2.37±0.12 | 632.8 | | |
| | [4] | -3.06±0.42 | 632.8 | | |
| | [5] | -2.56±0.13 | 514.5 | 290.15 | 760 |
| | [6] | -2.62±0.08 | 632.8 | 203-393 | |
| | [7] | -2.43±0.12 | 632.8 | 203-393 | |
| | [8] | -2.26±0.10 | 514.5 | | |
| | [15] | -2.17±0.21 | 1064 | | |
| | [14] | $-2.66\pm0.42_{(\pm0.12}{}^{\dagger}{}_{\pm0.40}{}^{\ddagger})$ | 1064 | 292.26-292.96 | 2-150 |
| | [§] | $-2.02\pm0.18_{(\pm0.16}{}^{\dagger}{}_{\pm0.08}{}^{\ddagger})$ | 1064 | 294.86-297.96 | 2-50 |
| $O_2$ | [5] | -2.52±0.06 | 514.5 | 290.15 | |
| (M = 12) | [6] | -2.52±0.06 | 632.8 | 200-400 | |
| | [9] | -2.56±0.04 | 632.8 | 298.6-463.7 | |
| | [17] | -2.29±0.08 | 1064 | 293.15 | |
| | [§] | $-1.79\pm0.35_{(\pm0.34}{}^{\dagger}{}_{\pm0.08}{}^{\ddagger})$ | 1064 | 294.86-297.96 | 0.5-30 |
| $Ar$ | [7] | 13.4±11.2 | 546.1 | 193.15-293.15 | |
| (M = 15) | [30] | 6.8±1.0 | 514.5 | 273.15 | |
| | [31] | 8.69±1.58 | 790 | 285 | 787.76 |
| | [§] | $4.31\pm0.38_{(\pm0.34}{}^{\dagger}{}_{\pm0.17}{}^{\ddagger})$ | 1064 | 294.86-297.96 | 50-295 |
| $Kr$ | [30] | 9.9±1.1 | 514.5 | 273.15 | |
| (M = 15) | [32] | 10.2±0.7 | 632.8 | 273.15 | |
| | [15] | 8.61±0.35 | 1064 | | |
| | [§] | $8.28\pm1.30_{(\pm1.26}{}^{\dagger}{}_{\pm0.32}{}^{\ddagger})$ | 1064 | 294.86-297.96 | 30-100 |
| $CO_2$ | [2] | -5.61±0.28 | 546.1 | 293.15 | |
| (M = 13) | [3] | -5.61±0.25 | 632.8 | 293.15 | |
| | [4] | -5.90±0.94 | 632.8 | 293.15 | |
| | [7] | -5.90±0.12 | 632.8 | 294.15 | |
| | [§] | $-4.22\pm0.31_{(\pm0.27}{}^{\dagger}{}_{\pm0.16}{}^{\ddagger})$ | 1064 | 294.86-297.96 | 2-50 |

§: present work; †: statistical uncertainty $\sigma_{\Delta nu\text{-}stat}$; ‡: systematic uncertainty $\sigma_{\Delta nu\text{-}sys}$.